\documentclass[iop]{emulateapj}

\newcommand\etal{et al.}
\newcommand\chandra{{\it Chandra}}
\newcommand\xmm{{\it XMM-Newton}}
\newcommand\sat{{\it INTEGRAL\/}}
\newcommand\psr{PSR J1101$-$6101}
\newcommand\psra{PSR J1105$-$6107}
\newcommand\src{IGR J11014$-$6103}
\newcommand\snr{MSH~11$-$6{\it 1}A} 
\newcommand\gsnr{G290.1$-$0.8}
\newcommand\per{62.8~ms}
\newcommand\pzero{$54\le P_0\le60$~ms}

\newcommand\effval{15.9235474(14)~s$^{-1}$}

\newcommand\fdotval{$(-2.17\pm0.13)\times10^{-12}$~s$^{-2}$}
\newcommand\edot{$\dot E=1.36\times10^{36}$ erg~s$^{-1}$}
\newcommand\edotval{$1.36\times10^{36}$ erg~s$^{-1}$}
\newcommand\edotdef{$\dot E=-4\pi^2 I f\dot f=1.36\times10^{36}$ erg~s$^{-1}$}
\newcommand\tauc{$\tau_c=116$~kyr}
\newcommand\taucval{116~kyr}
\newcommand\taucdef{$\tau_c\equiv|f/2\dot f|=116$~kyr}
\newcommand\bs{$B_s=7.4\times10^{11}$~G}
\newcommand\bsval{$7.4\times10^{11}$~G}
\newcommand\bsdef{$B_s=3.2\times10^{19}\,(P\dot P)^{1/2}$~G $=7.4\times10^{11}$~G}

\slugcomment{To Appear in The Astrophysical Journal Letters}

\shorttitle{X-ray Pulsations from \src}
\shortauthors{Halpern et al.}

\begin{document}

\title{Discovery of X-ray Pulsations from the \sat\ Source \src}

\author{
J.~P. Halpern\altaffilmark{1},
J.~A. Tomsick\altaffilmark{2},
E.~V. Gotthelf\altaffilmark{1},
F.~Camilo\altaffilmark{1},
C.-Y. Ng\altaffilmark{3},
A.~Bodaghee\altaffilmark{4},
J.~Rodriguez\altaffilmark{5},
S.~Chaty\altaffilmark{5,6},
F.~Rahoui\altaffilmark{7}
}

\altaffiltext{1}
{Columbia Astrophysics Laboratory, Columbia University, 550 West 120th Street,
New York NY, 10027, USA; jules@astro.columbia.edu}
\altaffiltext{2}
{Space Sciences Laboratory, 7 Gauss Way, University of California,
Berkeley, CA 94720-7450, USA}
\altaffiltext{3}
{Department of Physics, The University of Hong Kong, Pokfulam Road,
Hong Kong, China}
\altaffiltext{4}
{Georgia College \& State University, CBX 82, Milledgeville, GA 31061, USA}
\altaffiltext{5}
{Laboratoire AIM (UMR-E 9005 CEA/DSM-CNRS-Universit\'e Paris Diderot),
Irfu/Service d'Astrophysique, CEA-Saclay, F-91191 Gif-sur-Yvette Cedex,
France}
\altaffiltext{6}
{Institut Universitaire de France, 103 boulevard Saint-Michel, 75005 Paris,
France}
\altaffiltext{7}
{European Southern Observatory, Karl Schwarzschild-Strasse 2, D-85748
Garching bei M\"unchen, Germany}

\begin{abstract}
We report the discovery of \psr,
a 62.8~ms pulsar in \src, a hard X-ray source with
a jet and a cometary tail that strongly suggests it is moving away
from the center of the supernova remnant (SNR)
\snr\ at $v>1000$~km~s$^{-1}$.  Two \xmm\ observations
were obtained with the EPIC pn in small window mode, resulting
in the measurement of its spin-down luminosity \edot,
characteristic age \tauc, and surface magnetic field strength \bs.  
In comparison to $\tau_c$, the $10-30$~kyr age estimated for \snr\ suggests
that the pulsar was born in the SNR with initial period in the range \pzero.
\psr\ is the least energetic of the 15 rotation-powered pulsars detected by
\sat, and has a high efficiency of hard X-ray radiation and jet power.
We examine the shape of the cometary nebula in a \chandra\ image, 
which is roughly consistent with a bow shock at the velocity
inferred from the SNR age and the pulsar's $\dot E$.  However, its structure
differs in detail from the classic bow shock, and we explore possible
reasons for this.
\end{abstract}

\keywords{ISM: individual objects (\snr, \gsnr) ---
pulsars: individual (\psr, \psra) --- stars: neutron ---
X-rays: individual (\src)}

\section{Introduction}

\src\ was discovered as a hard X-ray ($20-100$\,keV) source in
\sat\ observations of the Galactic plane \citep{bir10}. 
\chandra\ and \xmm\ images show that
it has a complex X-ray morphology \citep{pav11,tom12,pav14},
consisting of a point source, a cometary pulsar wind nebula
(PWN) extending
$1.\!^{\prime}2$ northeast of the point source, an apparent
$\approx5.\!^{\prime}5$ long jet that is oriented perpendicular to
the PWN, and a faint counterjet.  The PWN points back to the
center of the supernova remnant \snr\ (=\gsnr), and its shape and
distance from the SNR suggest that the pulsar was born there and
kicked with high velocity \citep{tom12}.
The X-ray PWN is associated with the radio source
MGPS-2 J110149$-$610104 detected at 843\,MHz \citep{pav11}. 
Further mapping with the
Australia Telescope Compact Array at 2~GHz revealed that
the radio source has a bow-shock
morphology similar to that of the X-rays \citep{pav14}.

The highly collimated, $\approx5.\!^{\prime}5$ long X-ray jet most
likely parallels the rotation axis of the pulsar, which is therefore
oriented nearly perpendicular to the velocity vector.
With these properties, \src\ resembles the Guitar Nebula
associated with the high velocity pulsar PSR~B2224+65
\citep{hui07,hui12,joh10}.  The twisted jet (and faint counterjet)
of \src\ were modelled by \citet{pav14} as
a precessing, ballistic outflow emitting synchrotron radiation.

\snr\ is a mixed-morphology SNR whose centrally bright, thermal X-ray
emission observed by the {\it Advanced Satellite for Cosmology and
Astrophysics} ({\it ASCA\/}) was analyzed by \citet{sla02} using two
evolutionary models: thermal conduction, and cloudy ISM.
The results are that the SNR is $10-20$ kyr old and is at a
distance of $8-11$~kpc.  Since \psr\ is $11^{\prime}\!.9$ from the center
of \snr, the evolutionary model constraints imply a tangential velocity
of $v_{\perp}=2400$~km~s$^{-1}$ and 2900~km~s$^{-1}$ for the thermal
conduction and cloudy ISM models, respectively. (Age and distance
are correlated in these models.)  If this scenario is correct,
\psr\ would be the highest velocity pulsar known.
However, \citet{rey06} measured a smaller
distance of $7\pm1$~kpc to \snr\ from \ion{H}{1} 21~cm absorption.
Using \xmm\ and
\chandra\ data, \citet{gar12} derived an age range for
\snr\ of $10-30$~kyr.   With these revisions, the pulsar's kick velocity 
is still $>800$~km~s$^{-1}$, an exceptional value compared, e.g., to the
mean two-dimensional velocity of $307\pm47$~km~s$^{-1}$ for young
pulsars \citep{hob05}.

\section{Observations and Results}
\label{f:obs}

\begin{deluxetable*}{lccccccr}
\tabletypesize{\footnotesize}
\tablewidth{0pt}
\tablecaption{\xmm\ Timing Observations of \psr}
\tablehead{
\colhead{Instr/Mode} & \colhead{ObsID} &
\colhead{Date (UT)} & \colhead{Date (MJD)} & \colhead{Exp (s)} &
\colhead{Counts\tablenotemark{a}} &
\colhead{Frequency (Hz)\tablenotemark{b}} & \colhead{$Z_1^2$}
}
\startdata
EPIC-pn/SW & 0722600101 & 2013 July 21 & 56494.033 & 38000 & 2110 & 15.9235473(14) & 123.5 \\
EPIC-pn/SW & 0740880201 & 2014 June 8  & 56816.645 & 36476 & 1997 & 15.9234868(19) &  68.5
\enddata
\tablenotetext{a}{Background subtracted source counts in the 0.5--10 keV band
from a $15^{\prime\prime}$ radius aperture.}
\tablenotetext{b}{$1\sigma$ uncertainty in parentheses.}
\label{tab:log}
\end{deluxetable*}

\subsection{Pulsar Discovery and Timing}

We made two \xmm\ timing observations of \src\ separated by 322 days.
The EPIC pn CCD was operated in small window mode, which has a 5.7\,ms
sampling time.  The two EPIC MOS detectors were used in full frame
mode to image the entire PWN and jet.  This Paper reports only the
timing results from the pn CCD.  Table~\ref{tab:log}
is a log of the pn observations, indicating the net useable exposure
time (elapsed, i.e., not reduced for dead-time), and the measured signal.
The first observation was 44~ks long, but its final 6~ks were contaminated
by high radiation background near perigee; thus, we use only the first 38~ks.
The second observation was a clean 36.5~ks, and required no filtering.

\begin{figure}[b]
\begin{center}
\includegraphics[width=0.95\linewidth,angle=270]{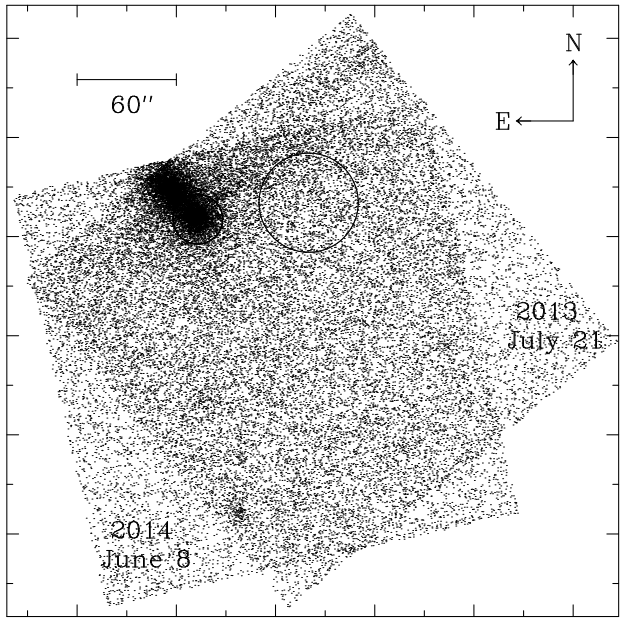}
\end{center}
\caption{
The two EPIC pn small window ($4.\!^{\prime}3\times4.\!^{\prime}3$)
images of \src\ listed in Table~\ref{tab:log}.
Extraction regions are the small circle ($15^{\prime\prime}$ radius)
for \psr\ and the large circle ($30^{\prime\prime}$ radius) for background.
}
\label{fig:pnimage}
\end{figure}

\begin{deluxetable}{ll}
\tablecaption{Timing Parameters for \psr}
\tablewidth{0pt}
\tablehead{
\colhead{Parameter}   &
\colhead{Value}
}
\startdata                                       
R.A. (J2000.0)\tablenotemark{a}   & $11^{\rm h}01^{\rm m}44.\!^{\rm s}96$ \\
Decl. (J2000.0)\tablenotemark{a}  & $-61^{\circ}01^{\prime}39.\!^{\prime\prime}6$ \\
Epoch (MJD TDB)\tablenotemark{b}                 & 56494.00000012   \\
Frequency\tablenotemark{c}, $f$                  & \effval\         \\
Frequency derivative\tablenotemark{c}, $\dot f$  & \fdotval\        \\
Period\tablenotemark{c}, $P$                     & 0.062800077(6) s \\
Period derivative\tablenotemark{c}, $\dot P$     & $(8.56\pm0.51)\times10^{-15}$ \\
Range of dates (MJD)                             & 56494--56817     \\
Spin-down luminosity, $\dot E$                   & \edotval\        \\
Characteristic age, $\tau_c$                     & \taucval\        \\
Surface dipole magnetic field, $B_s$             & \bsval\
\enddata
\tablenotetext{a}{\chandra\ position from \citet{tom12}.}
\tablenotetext{b}{Epoch of phase zero in Figure~\ref{fig:pulse}.}
\tablenotetext{c}{$1\sigma$ uncertainty in parentheses.}
\label{tab:ephem}
\end{deluxetable}

\begin{figure}[b]
\begin{center}
\includegraphics[width=0.7\linewidth,angle=270]{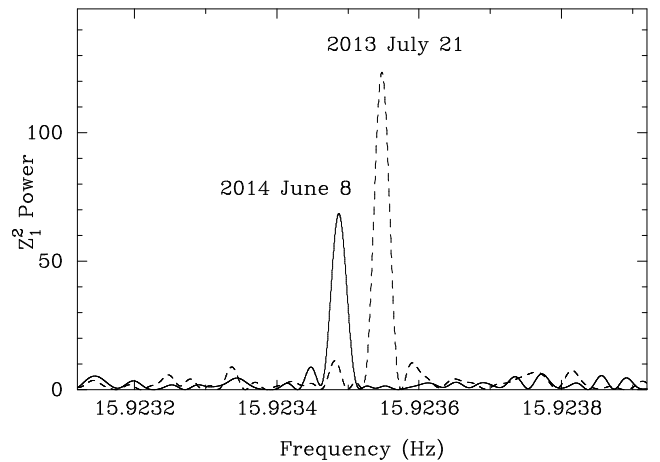}
\end{center}
\caption{
$Z_1^2$ periodograms from the two \xmm\ timing observations
listed in Table~\ref{tab:log}.  The change in frequency
corresponds to $\dot f=$\fdotval.  
}
\label{fig:zplot}
\end{figure}

Events in the 0.5$-$10\,keV band were selected from a circle of radius
$15^{\prime\prime}$ around the point source.  This choice was a compromise
between maximizing the counts extracted from the pulsar and minimizing
contamination from the adjacent bow-shock nebula and jet.
Figure~\ref{fig:pnimage} shows the two images superposed, with the
extraction circle for the pulsar and another circle 
used for background estimation.  The photon arrival times were transformed
to barycentric dynamical time using the \chandra\ measured position
of the point source \citep{tom12}.  The $Z_1^2$ test (Rayleigh test,
\citealt{str80,buc83}) was used to search for pulsations, and a single,
highly significant peak was found in each observation at a period of \per.
The $Z_1^2$ periodograms are shown in Figure~\ref{fig:zplot},
where the peak values are 123.5 and 68.5.
Noise power $S$ is distributed as $0.5\,e^{-S/2}$,
and the number of independent trials in a search to the Nyquist frequency
is $\approx3\times10^6$.  This leads to negligible probabilities
of $5\times10^{-21}$ and $4\times10^{-9}$, respectively, that the two
detections are false.  One-sigma
uncertainties in the peak frequencies were estimated from the range
corresponding to $Z_1^2({\rm max})-1$ around the peak. The significant change
in frequency results in a measurement of its derivative with 6\% precision.
Further examination of the radio timing data reported by
\citet{tom12} does not reveal a signal; thus, \psr\ remains radio quiet
to the same limit derived in \citet{tom12}.

Table~\ref{tab:ephem} lists the derived dipole spin-down parameters
of \psr, including the spin-down luminosity \edotdef, the characteristic
age \taucdef, and the surface dipole magnetic field strength \bsdef.  An
important caveat is the possibility that an intervening
glitch may have biassed the measurement of $\dot f$.  The fractional
change in frequency over 322 days is $\Delta f/f=-3.8\times10^{-6}$.
This can be compared to the largest glitches in the Vela pulsar,
which have $\Delta f/f\sim2\times10^{-6}$ and a mean recurrence
time of $\approx3$~yr \citep{esp11}.  If \psr\ glitched between
the epochs of our observations, it is possible that its spin-down
rate has been underestimated by as much as $\sim50\%$.
However, \psr\ is not likely to be as active as the Vela pulsar,
which is a uniquely strong and frequent glitcher.
In any case, our main conclusions would not be changed by
a $\sim50\%$ revision in age or spin-down power.

The difference in peak $Z_1^2$ values of the two observations,
after scaling for exposure time, is not great enough to claim 
that the pulsed fraction has changed.
The variance in measured power as a function of intrinsic power was
treated by \citet{gro75}, and is summarized in Figure~1 of that paper
(with the difference that Groth's power is actually our $Z_1^2/2$).
The figure shows, for example, that if the true power is $Z_1^2=100$,
then there is a 16\% chance that the measured power will be $>120$,
and a 5\% chance that it will be $<68$.

\begin{figure}
\begin{center}
\includegraphics[width=0.75\linewidth,angle=0]{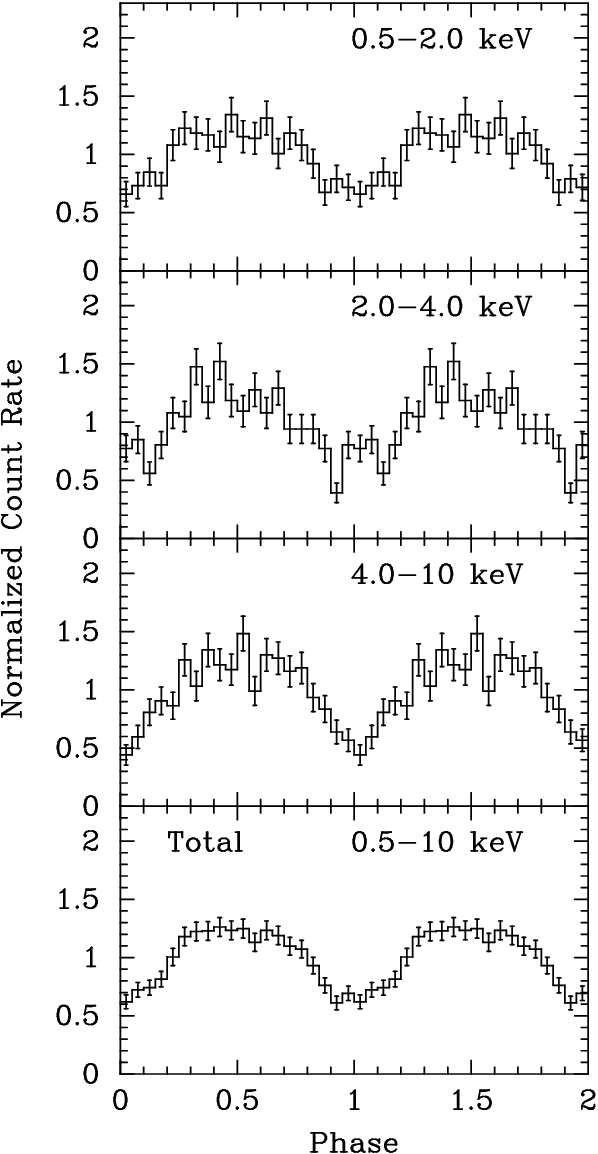}
\end{center}
\caption{
Energy-dependent pulse profiles of \psr\
from the two \xmm\ timing observations combined.
They are background subtracted and normalized to 1.
The phase between the two observations
was adjusted to align them.
}
\label{fig:pulse}
\end{figure}

We used the timing parameters to combine the pulse
profiles of the two observations, adjusting their
relative phase to maximize $Z_1^2$ in the combined data.
The folded light curves in a range of energies between 0.5 and 10~keV
are shown in Figure~\ref{fig:pulse}, where they
are background-subtracted and normalized to 1 in
each energy band.  The single-peaked
pulse shape and its phasing appears to be independent of
energy, while the pulsed fraction increases from $\approx 35\%$
at the lowest energy to $\approx50\%$ at the highest.
However, the pulsed fractions shown in Figure~\ref{fig:pulse} are
definitely lower limits, as the source extraction circle
includes an unknown number of counts from the PWN
to the northeast and, to a lesser extent, from the jet, while the
circle used for background subtraction (Figure~\ref{fig:pnimage})
does not correct for this contamination.
Although the effect is difficult to quantify, it probably
accounts for the apparent increase in pulsed fraction with
energy in Figure~\ref{fig:pulse}, as the PWN
has a softer spectrum than the pulsar
($\Gamma_{\rm PWN}=1.9\pm0.1$, $\Gamma_{\rm PSR}=1.1\pm0.2$; \citealt{pav14}). 
The intrinsic pulsed fraction is therefore likely to be $\ge50\%$
at all energies.  In support of this interpretation, we find that when
we decrease the radius of the extraction aperture from $15^{\prime\prime}$
to $10^{\prime\prime}$, the pulsed fraction becomes
$\approx50\%$ at all energies. 

\subsection{Bow-Shock Fitting}

\begin{figure}
\begin{center}
\includegraphics[width=0.95\linewidth,angle=270]{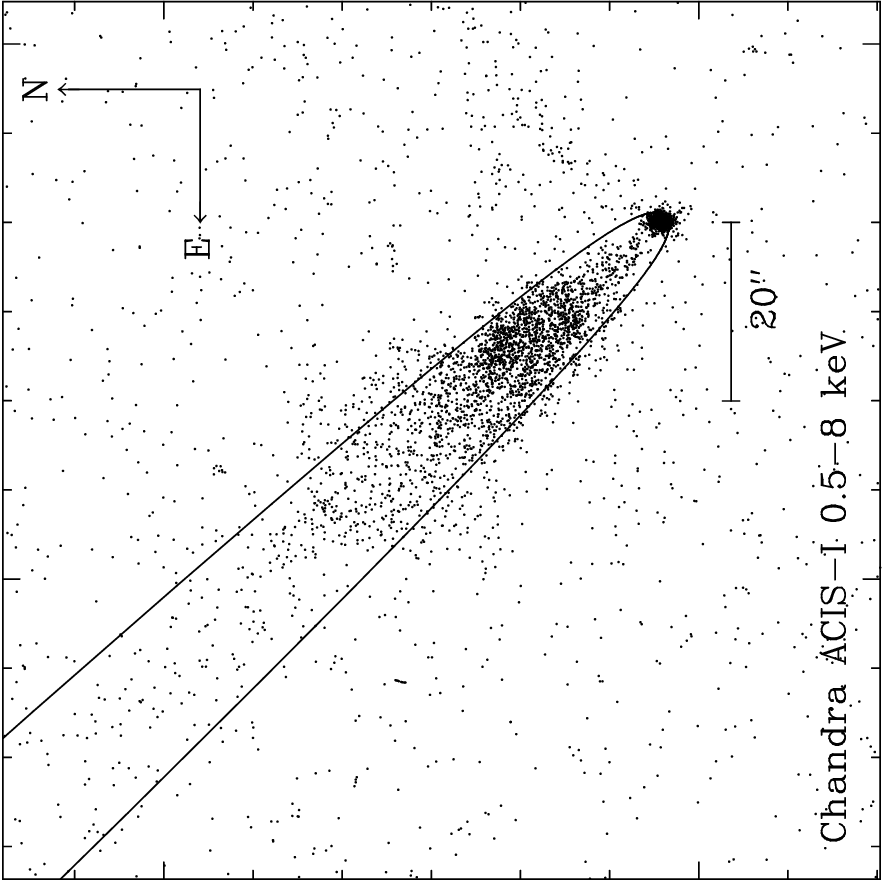}
\end{center}
\caption{
\chandra\ image of \psr\ and its PWN
from the 49.4~ks observation of \citet{pav14}.
Superposed is the \citet{wil96} equation of the contact
discontinuity between the shocked pulsar wind and the shocked ISM,
fitted by eye.  The parameters are the position
angle of motion, $223^{\circ}$, and the radius
of the apex, or stagnation point, which is
$r_0=7.9\times10^{16}\,d_7$~cm ($0.\!^{\prime\prime}75$)
from the pulsar.  The inclination angle $i$ of the velocity vector
with respect to the plane of the sky is assumed to be $0^{\circ}$.
}
\label{fig:chandra}
\end{figure}

Knowing the spin-down power of \psr, we reexamine the
structure of its apparent bow-shock nebula to obtain an
independent estimate of the space velocity of the pulsar.
For this purpose, the higher resolution of \chandra\ is
more useful than \xmm.  Our analysis here follows and extends
that of \citet{tom12}, who used a 5~ks \chandra\
observation (ObsID 12420) with the Advanced CCD Imaging
Spectrometer.   Here, we revisit the 49.4~ks \chandra\
ACIS-I observation (ObsID 13787) that was presented by \citet{pav14}.
Figure~\ref{fig:chandra} shows the region of this image containing
the pulsar and PWN, with each photon in the $0.5-8$~keV band
indicated by a dot.   Because the pulsar was located only
$0.\!^{\prime}$8 from the optical axis, the spatial resolution
for the pulsar and its immediate surroundings is nearly optimal.

For the case of an isotropic wind from a star moving
supersonically through a uniform ISM, \citet{wil96}
derived an analytic expression for the surface of
contact discontinuity between the shocked pulsar wind and the
shocked ISM using momentum conservation,
$r(\theta)=r_0\,{\rm csc}\,\theta\,[3\,(1-\theta\,{\rm cot}\,\theta)]^{1/2}$,
where $\theta$ is the polar angle with respect to direction of motion,
$r(\theta)$ is the distance of the surface from the star, and $r_0$
is the stagnation radius, the distance of the apex of the surface
from the pulsar.  The shape of the contact discontinuity is thus
parameterized entirely in terms of $r_0$, which in turn is determined
in this case by the pulsar wind power, assumed to be $\approx\dot E$,
the velocity $v$ of the pulsar, and the ambient density $\rho$ of the
ISM, using momentum balance:
$$r_0 = \left(\dot E \over 4\pi\,\rho\,v^2\,c\right)^{1/2}.\eqno{(1)}$$

The assumption we make in graphing the model curve
in Figure~\ref{fig:chandra}
is that the X-ray emission comes from the shocked pulsar wind,
which is bounded by the contact discontinuity and the termination
shock interior to it.  Therefore, $r_0$ was chosen by eye so
that the curve surrounds the bulk of the emission trailing the pulsar.
It was assumed that the pulsar is moving nearly in the plane
of the sky ($i=0^{\circ}$), both because of
the narrow opening angle of the nebula, and the evidently
large tangential velocity.  The parameters of the curve
are $r_0=7.9\times10^{16}\,d_7$~cm ($0.\!^{\prime\prime}75$),
where $d_7$ is the distance in units of 7~kpc,
and the position angle of the motion,
$223^{\circ}$.  Similar results were obtained
by \citet{tom12}.  Now substituting the values of $r_0$ and
$\dot E$, Equation (1) is reduced to
$v_{\perp}=500\,d_{7}^{-1}\,n_{0.1}^{-1/2}$~km~s$^{-1}$,
where $n_{0.1}$ is the ISM hydrogen density in units of 0.1~cm$^{-3}$. 
This velocity is $\sim2-4$ times less than the estimates
from the SNR age, but they can be reconciled if the local
density is $\sim0.01$~cm$^{-3}$.  The result is largely unaffected
by the unknown angle $i$, because any inclination of the
model would broaden the apparent opening angle of the bow shock,
which would then have to be reduced by decreasing $r_0$, thus
increasing $v$.

However, before giving this analysis too much credence,
note that there are at least two discrepancies
between the simple model and the detailed properties of
the data.  First, there is no evidence of
emission from the apex of the shocked wind, which
theoretically should be the brightest part of the nebula.
In fact, there is no excess emission within
$\approx3^{\prime\prime}$ of the pulsar, a zone that
is entirely consistent with a single point source
\citep{tom12,pav14}.  Second, the faint X-ray emission
just behind the pulsar does not in fact follow the model
curve, but is confined to a narrower cone, while the
brightest regions filling the curve are
$13^{\prime\prime}-26^{\prime\prime}$ behind the pulsar.
The appearance is of a diverging flow getting
brighter with distance from the pulsar, rather
than a collimated one that is fading.
Similar phenomena have been noted in the X-ray
images of other pulsar tails; the
possible implications will be discussed in
Section~\ref{f:structure}.

\section{Discussion}

\subsection{Distance and Associations}

Various estimates of the distance to \snr\ were
reviewed by \citet{fil05}, who concluded from
their own CO maps that $d=7-8$~kpc, in agreement
with 6.9~kpc from optical
emission-line velocities \citep{ros96},
and $7\pm1$~kpc from \ion{H}{1} 21~cm absorption \citep{rey06}.
These are all kinematic distances, unlike the
X-ray modelling of \citet{sla02}, who derived
$d=8-11$~kpc.  We have adopted 7~kpc as the most likely distance.

Since the characteristic age of \psr\ is greater than all estimates
of the age of \snr, we can assume that it was born in the SNR
and estimate its birth period $P_0$ from the relation
$$T = {P \over (n-1)\dot P}\left[1-\left(P_0 \over P\right)^{n-1}\right],$$ 
where $n\equiv f\ddot f/\dot f^2$ is the braking index.  Most pulsars
have $2<n<3$ \citep{liv07}.  For this range, and letting
$T$ be the $10-30$~kyr age of the SNR, we find \pzero.

\psr\ should not be confused with \psra, a 63.2~ms pulsar with
$\dot E=2.5\times10^{36}$ erg~s$^{-1}$ and $\tau_c=63$~kyr that
is $23^{\prime}$ southeast of \snr.  \citet{kas97} considered the
possibility that \psra\ was born in \snr, although
\psr\ is now a more compelling association.
The dispersion measure of 271~pc~cm$^{-3}$ to \psra\
converts to a distance of 5.0~kpc according to the
Galactic electron density model of \citet{cor02}.
The corresponding free-electron column of
$N_{\rm e}=8.4\times10^{20}$~cm$^{-2}$,
assuming a typical ionized fraction of 0.1 \citep{he13},
is accompanied by a neutral column of
$N_{\rm H}\approx8.4\times10^{21}$~cm$^{-2}$, 
which is consistent with the X-ray measured
$N_{\rm H}=8\times10^{21}$~cm$^{-2}$ to \src\ \citep{tom12,pav14}.
X-ray measurements of $N_{\rm H}$ to \snr\ are somewhat
contradictory, ranging from $(4.3-6.2)\times10^{21}$~cm$^{-2}$
\citep{gar12} to $(1.3\pm0.1)\times10^{22}$~cm$^{-2}$ \citep{sla02}.
Allowing for this ambiguity, all three objects are probably
consistent with being at the same distance.

\psra\ is not detected in X-rays.  Using archival \chandra\ observations
totaling 23.7~ks (ObsIDs 2780 and 4380), we set a $3\sigma$ upper limit of
$8\times10^{-15}$ erg~cm$^{-2}$~s$^{-1}$ on its $2-10$~keV
flux, corresponding to $L_X/\dot E<2\times10^{-5}\,d_7^2$.  This
is close to the minimum of the distribution of similarly aged pulsars
\citep{kar08}.
A previously claimed detection of this pulsar using {\it ASCA\/}
\citep{got98} may have instead detected a neighboring star that
is present in the \chandra\ images.

\subsection{Energetics}

With \edot, \psr\ is the least energetic of the 15 rotation-powered
pulsars detected by \sat\ (for the full set see 
\citealt{mat09,ren10,got11,hal12}).
These are among the most energetic pulsars, 
comprising half of all those known with
$\dot E\ge3.7\times10^{36}$\,erg~s$^{-1}$, the latter value
belonging to PSR B1951+32, which has
a characteristic age of 107~kyr and was previously the
least energetic of the \sat\ pulsars.

The $20-100$~keV flux of \src\ is $8.7\times10^{-12}$ erg~cm$^{-2}$~s$^{-1}$
\citep{bir10}, corresponding to a luminosity of
$5.1\times10^{34}\,d_7^2$~erg~s$^{-1}$
that is 4\% of the spin-down luminosity of \psr.  This
exceeds the combined $2-10$ keV flux of the pulsar, the PWN, and the jet
as measured by \chandra, which total
$1.8\times10^{-12}$ erg~cm$^{-2}$~s$^{-1}$ \citep{pav14},
or 0.8\% of the spin-down flux.  The flat spectrum of the pulsar must
extend into the hard X-rays, where it is responsible for most of
the $20-100$~keV flux.  Using equipartition arguments, \citet{pav14}
estimated that a minimum power of $2\times10^{35}$ erg~s$^{-1}$
is needed for the jet, which is $14\%$ of the spin-down luminosity.
This is important evidence that a large part of a pulsar's
spin-down power can be focussed into a narrow polar jet,
a fraction therefore not available to power a bow shock.

\subsection{Structure of the PWN}
\label{f:structure}

The absence of X-rays from the head of the putative
bow shock is the principal challenge to the model in which
the termination shock of the pulsar is the cause of the PWN
emission.  A dark region between the termination shock
and the contact discontinuity is
difficult to understand in the context of shock acceleration.
For reasonable values of the magnetic field strength
($B\sim10^{-4}$~G) and X-ray emitting electron energy ($E\sim10^{13}$~eV),
the gyroradius, $r_g = 3\times10^{14}\,E_{13}\,B_{-4}^{-1}$~cm,
is much smaller than the stagnation radius,
$r_0=7.9\times10^{16}\,d_7$~cm. So the particles
should be easily confined and accelerated.

The cavity {\it interior} to the termination shock should
be dark, so that any emission from around it should be limb brightened.
The termination shock will close behind the pulsar at
a distance $r_1$, which is larger than its forward radius, $\approx r_0$.
In analytic and numerical models at low Mach number $\cal M$,
the relation between these radii is $r_1/r_0\approx\gamma^{1/2}\,{\cal M}$,
where $\gamma$ is the adiabatic index of the ambient medium,
usually 5/3 \citep{buc02}.  However, numerical models at high
$\cal M$ show that this ratio saturates at about 5 \citep{gae04}.
While an area $\approx12^{\prime\prime}$ long behind \psr\ is
relatively dim in X-rays, the emission there looks like a
narrow cone rather than the expected limb-brightened bow shock.
Nevertheless, if we ignore this detail and assume that this
is the region bounded by the termination shock, then
$r_1/r_0\approx12^{\prime\prime}/0.\!^{\prime\prime}75$
and $\cal M$$\,\approx12.4$.
The sound speed in the warm (8000~K) phase of the
ISM is $\approx13$~km~s$^{-1}$, which then implies
a pulsar velocity of only $\approx165$~km~s$^{-1}$, at odds with the
other estimates.  Only if the ISM is hot ($\sim10^6$~K)
do we get $v\approx1900$~km~s$^{-1}$.  But this would require
reducing the ambient density drastically to have a
reasonable pressure, which would be inconsistent
with the results from Equation (1).

\citet{gae04} suggested that the ``tongue''of emission
just behind PSR~J1747$-$2958 (the Mouse) and others
represents the surface of the termination shock.  But these are
bright regions, which contradicts the theory that there should
be no emission interior to the termination shock. In the case of
\psr, this region is at least underluminous, although not
limb-brightened. Another system like \psr\ in which trailing
emission brightens with distance from the pulsar is PSR~J0357+3205
\citep{del11,del13,mar13}.  The difficulties in modeling
that tail as a synchrotron emitting bow shock led the
authors to propose shocked-heated bremsstrahlung emission
instead.  But that model requires a hot ISM phase 
with an extraordinarily large pressure.

In several pulsar tails, radio and X-ray brightness are
anticorrelated, with the radio
increasing with distance from the pulsar \citep{ng10}.
This is the case for \psr\ as well \citep{pav14}.
However, it is not clear if this phenomenon relates to why the
region closest to \psr\ is underluminous in both radio and X-ray.

Considering that the spin axis of \psr\ may be orthogonal
to its velocity vector, with a large fraction of the
spin-down power going into the jet, one may ask if
the remaining wind is primarily polar or equatorial, and
how that would affect the structure of the PWN.  However,
numerical models with anisotropic pulsar winds, including an
equatorial one, do not significantly change the shape of the
termination shock \citep{vig07}.  So far, no model within
the framework of ideal MHD appears to explain the features
of our data and others.

An alternative model \citep{rom05} in which particles are
accelerated by magnetic reconnection outside the speed-of-light
cylinder results in a fast ``magnetotail'' behind the pulsar,
which may contain a large fraction of the energy of the pulsar
wind and extend to large distances.  An interesting feature of
this model is the flared ``trumpet'' shape of the magnetotail
(Figure 4 of \citealt{rom05}),
which does in fact resembles the PWN of \psr.  An approximation
for the radiation length of the magnetotail is
$$r\approx15\,P^2\,\left(B_s\over 10^{12}\,{\rm G}\right)^{-1}\,\left(n\over
{\rm cm}^{-3}\right)^{-1}\left(v\over 1000\,{\rm km\,s}^{-1}\right)^{-2}
\ {\rm pc}.$$
Assuming $n=0.03$~cm$^{-3}$ and $v=1000$~km~s$^{-1}$, this reduces to
$r\approx 2.7$~pc ($1.\!^{\prime}3$), the actual length of the PWN.
However, the authors only investigated the case
in which the magnetic axis, the rotation axis, and the velocity
are all parallel, while there is good reason to believe
that the rotation axis of \psr\ is nearly orthogonal to its
velocity because of the orientation of the jet.  It is not clear
if a more realistic geometry would generate undesirable,
non-axisymmetric features.

\section{Conclusions}

We discovered the \per\ pulsar \psr\ in \src.  Its spin-down
luminosity of \edotval\ is the lowest among the 15 rotation-powered
pulsars detected by \sat, and an order-of-magnitude less than what
was anticipated from the X-ray luminosity of \psr\ \citep{pav14}.
However, there is a large scatter among
pulsars in efficiency of X-ray emission.
Its \taucval\ characteristic age is consistent with
an origin in \snr\ for any reasonable value of the braking
index or SNR age, with its birth period
close to its present period. 

The velocity of the pulsar inferred from fitting
the shape of its cometary nebula is compatible with
estimates of $800-2400$~km~s$^{-1}$ from the SNR age and distance,
if the density of the ambient ISM is $<0.1$~cm$^{-3}$.
The density should be this low if \psr\ is within a
cavity blown by previous stellar winds or supernovae. 
Because the structure of the nebula differs in important details
from a basic bow-shock geometry, we are not secure in making
quantitative estimates of velocity and density from such
a simple model. Interestingly, however, an alternative magnetotail
theory would require similar velocity and density.


\begin{thebibliography}{}

\bibitem[Bird \etal(2010)]{bir10}
Bird, A. J., Bazzano, A., Bassani, L., \etal\ 2010, \apjs, 186, 1

\bibitem[Bucciantini(2002)]{buc02}
Bucciantini, N. 2002, \aap, 387, 1066

\bibitem[Buccheri \etal(1983)]{buc83}
Buccheri, R., Bennett, K., Bignami, G. F., \etal\ 1983, \aap, 128, 245

\bibitem[Cordes \& Lazio(2002)]{cor02}
Cordes, J. M., \& Lazio, T. J. W. 2002, arXiv:astro-ph/0207156

\bibitem[De Luca \etal(2011)]{del11}
De Luca, A., Marelli, M., Mignani, R. P., \etal\ 2011, \apj, 733, 104

\bibitem[De Luca \etal(2013)]{del13}
De Luca, A., Mignani, R. P., Marelli, M., \etal\ 2013, ApJL, 765, L19

\bibitem[Espinoza \etal(2011)]{esp11}
Espinoza, C. M., Lyne, A., Stappers, B. W., \& Kramer, M. 2011,
\mnras, 414, 1679

\bibitem[Filipovi\'c \etal(2005)]{fil05}
Filipovi\'c, M. D., Payne, J. L., \& Jones, P. A. 2005, Serb. Astron. J.,
170, 47

\bibitem[Gaensler \etal(2004)]{gae04}
Gaensler, B. M., van der Swaluw, E., Camilo, F., \etal\ 2004, \apj, 616, 383

\bibitem[Garcia et al.(2012)]{gar12}
Garcia, F., Combi, J. A., Albacete-Colombo, J. F., et al. 2012, \aap, 546, A91

\bibitem[Gotthelf \etal(2011)]{got11}
Gotthelf, E. V., Halpern, J. P., Terrier, R., \& Mattana, F. 2011,
ApJL, 729, L16

\bibitem[Gotthelf \& Kaspi(1998)]{got98}
Gotthelf, E. V., \& Kaspi, V. M. 1998, ApJL, 497, L29

\bibitem[Groth(1975)]{gro75}
Groth, E. J. 1975, \apjs, 29, 285

\bibitem[Halpern \etal(2012)]{hal12}
Halpern, J. P., Gotthelf, E. V., \& Camilo, F. 2012, ApJL, 753, L14

\bibitem[He \etal(2013)]{he13}
He, C., Ng, C.-Y., \& Kaspi, V. M. 2013, \apj, 768, 64

\bibitem[Hobbs \etal(2005)]{hob05}
Hobbs, G., Lorimer, D.~R., Lyne, A.~G., \& Kramer, M. 2005,
\mnras, 360, 974

\bibitem[Hui \& Becker(2007)]{hui07}
Hui, C. Y., \& Becker, W. 2007, \aap, 467, 1209

\bibitem[Hui \etal(2012)]{hui12}
Hui, C. Y., Huang, R. H. H., Trepl, L., \etal\ 2012, \apj, 747, 74

\bibitem[Johnson \& Wang(2010)]{joh10}
Johnson, S. P., \& Wang, Q. D. 2010, \mnras, 408, 1216

\bibitem[Kargaltsev \& Pavlov(2008)]{kar08}
Kargaltsev, O., \& Pavlov, G. G. 2008, 
AIP Conf. Proc. 983, 40 Years of Pulsars:
Millisecond Pulsars, Magnetars and More.
ed. C. Bassa, Z. Wang, A. Cumming, \& V. M. Kaspi (Melville, NY: AIP), 171

\bibitem[Kaspi \etal(1997)]{kas97}
Kaspi, V., Bailes, M., Manchester, R. N., \etal\ 1997, \apj, 485, 820

\bibitem[Livingstone \etal(2007)]{liv07}
Livingstone, M. A., Kaspi, V. M., Gavriil, F. P., \etal\ 2007, Ap\&SS, 308, 317

\bibitem[Marelli \etal(2013)]{mar13}
Marelli, M., De Luca, A., Salvetti, D., \etal\ 2013, \apj, 765, 36

\bibitem[Mattana \etal(2009)]{mat09}
Mattana, F., G\"otz, D., Terrier, R., Renaud, M., \& Falanga, M. 2009,
AIP Conf. Proc. 1126, Simbol-X: Focusing on the Hard X-ray Universe,
ed. J. Rodriguez \& P. Forrando (Melville, NY: AIP), 259 

\bibitem[Ng \etal(2010)]{ng10}
Ng, C.-Y., Gaensler, B. M., Chatterjee, S., \& Johnston, S. 2010,
\apj, 712, 596

\bibitem[Pavan \etal(2014)]{pav14}
Pavan, L., Bordas, P., P\"uhlhofer, G., \etal\ 2014, 562, A122

\bibitem[Pavan \etal(2011)]{pav11}
Pavan, L., Bozzo, E., P\"uhlhofer, G., \etal\ 2011, \aap, 533, A74

\bibitem[Renaud \etal(2010)]{ren10}
Renaud, M., Marandon, V., Gotthelf, E. V., \etal\ 2010, \apj, 716, 663 

\bibitem[Reynoso et al.(2006)]{rey06}
Reynoso, E. M., Johnston, S., Green, A. J., et al. 2006, \mnras, 369, 416

\bibitem[Romanova \etal(2005)]{rom05}
Romanova, M. M., Chulsky, G. A., \& Lovelace, R. V. E. 2005, ApJ, 630, 1020 

\bibitem[Rosado \etal(1996)]{ros96}
Rosado, M., Ambrocio-Cruz, P., Le Coarer, E., \& Marcelin, M. 1996,
\aap, 315, 243

\bibitem[Slane \etal(2002)]{sla02}
Slane, P., Smith, R.~K., Hughes, J.~P., \& Petre, R. 2002, \apj, 564, 284

\bibitem[Strutt(1880)]{str80}
Strutt, J. W. 1880, PMag, 10, 73

\bibitem[Tomsick \etal(2012)]{tom12}
Tomsick, J.~A., Bodaghee, A., Rodriguez, J., \etal\ 2012, \apjl, 750, L39

\bibitem[Vigelius \etal(2007)]{vig07}
Vigelius, M., Melatos, A., Chatterjee, S., Gaensler, B. M.,
\& Ghavamian, P. 2007, MNRAS, 374, 793 

\bibitem[Wilkin(1996)]{wil96}
Wilkin, F. P. 1996, ApJL, 459, L31

\end{thebibliography}
\end{document}